\documentclass[12pt]{spieman}  
\usepackage{amsmath,amsfonts,amssymb}
\usepackage{graphicx}
\usepackage{setspace}
\usepackage{tocloft}
\usepackage{lineno}

\title{Transmission grating arrays for the X-ray spectrometer on Arcus Probe}

\author[a,*]{Ralf K. Heilmann}
\author[b]{Alexander R. Bruccoleri}
\author[c]{James A. Gregory}
\author[d]{Eric M. Gullikson}
\author[e]{Hans Moritz G\"unther}
\author[f]{Edward Hertz}
\author[c]{Renee D. Lambert}
\author[c]{Douglas J. Young}
\author[a]{Mark L. Schattenburg}
\affil[a]{Space Nanotechnology Laboratory, MIT Kavli Institute for Astrophysics and Space Research, Massachusetts Institute of Technology, Cambridge, MA 02139, USA}
\affil[b]{Izentis, LLC, Cambridge, MA 02139, USA}
\affil[c]{MIT Lincoln Laboratory, Lexington, MA 02421, USA}
\affil[d]{Lawrence Berkeley National Laboratory, Berkeley, CA 94720, USA}
\affil[e]{MIT Kavli Institute for Astrophysics and Space Research, Massachusetts Institute of Technology, Cambridge, MA 02139, USA}
\affil[f]{Center for Astrophysics, Harvard-Smithsonian Astrophysical Observatory, Cambridge, MA 02138, USA}

\cftpagenumbersoff{figure}
\cftpagenumbersoff{table} 
\begin{document} 
\maketitle

\begin{abstract}
The Arcus Probe mission concept has been submitted as an Astrophysics Probe Explorer candidate.  It features two co-aligned high-resolution grating spectrometers: one for the soft x-ray band and one for the far UV.  Together, these instruments can provide unprecedented performance to address important key questions about the structure and dynamics of our universe across a large range of length scales.  The X-ray Spectrometer (XRS) consists of four parallel optical channels, each featuring an x-ray telescope with a fixed array of 216 lightweight, high-efficiency blazed transmission gratings, and two CCD readout arrays.  Average spectral resolving power $\lambda/\Delta \lambda > 2,500$ ($\sim 3500$ expected) across the 12-50 \AA \ band and combined effective area $> 350$ cm$^2$ ($> 470$ cm$^2$ expected) near OVII wavelengths are predicted, based on the measured x-ray performance of spectrometer prototypes and detailed ray trace modeling.  We describe the optical and structural design of the grating arrays, from the macroscopic grating petals to the nanoscale gratings bars, grating fabrication, alignment, and x-ray testing.  Recent x-ray diffraction efficiency results from chemically thinned grating bars are presented and show performance above mission assumptions.
\end{abstract}


{\noindent \footnotesize\textbf{*}Corresponding author: Ralf K. Heilmann,  \linkable{ralf@space.mit.edu} }

\begin{spacing}{2}   

\section{Introduction}
\label{sect:intro}  

In our quest to understand the structure of the universe around us we cannot be satisfied with measuring how things are today, but we also must understand the dynamics of our universe, how our Milky Way and other galaxies formed and aggregated, and how they will develop into the future.  Images and low resolution spectra do not provide the required precision data that informs us about the kinetics and temperatures of baryons that underpin not only much of the visible universe, but also trace the small and large-scale structure and dynamics of its hottest components as highly ionized plasma.  Emission and absorption in these plasmas occurs predominantly in the soft x-ray and FUV bands, calling for high-resolution spectroscopy at these short wavelengths.  To quote the Astro2020 Decadal Review:\cite{Astro2020}
``Astronomy became astrophysics with the first spectrum. Spectroscopy determines compositions, magnetic field strength, space motion, rotation, multiplicity, planetary companions, surface structure, and other important physical traits...In the next decade, spectroscopy will be the dominant discovery tool for astronomy.”

Arcus Probe will carry the required state-of-the-art instruments that can deliver the desired spectra in much higher resolution and much shorter time than existing, aging x-ray observatories.
Details of the main science cases for the Arcus Probe Mission can be found in Ref.~\citeonline{Arcus2023}.

\section{The Arcus Probe X-ray Spectrometer (XRS)}

The X-ray Spectrometer (XRS) instrument consists of four parallel, almost identical optical channels (OC).  In each channel an array of grazing-incidence silicon pore optics (SPO),\cite{SPO_SPIE2023} with the layout of a sub-apertured version of the NewAthena design,\cite{NewAthenaSPO-SPIE2023} forms an x-ray telescope with 12 m focal length.  Sub-aperturing in azimuth for grazing-incidence mirrors leads to an anisotropic telescope point-spread function (PSF) that is much narrower in one dimension than a mirror array that is fully populated in azimuth.\cite{Cash1991,SPIE2010}. Just downstream of the x-ray mirrors follows a grating petal, carrying an array of 216 critical-angle transmission (CAT) gratings that covers the mirror array.  CAT gratings are blazed transmission gratings that disperse soft x rays into several diffraction orders, with the strongest orders on just one side of the image near twice the blaze angle of the gratings.  The gratings are aligned with their dispersion axes parallel to the narrow direction of the PSF, minimizing the spectrometer line-spread function (LSF) in the dispersion direction. Some soft and most harder x rays are not diffracted and form an image at the telescope focus, where x-ray CCDs provide energy resolution.\cite{GrantJATIS2024}  Effective area for harder x rays in $0^{\rm {th}}$ order peaks around 1400 cm$^2$ near 1.8 keV, and is still above 300 cm$^2$ at 6.7 keV (summed over all four OCs).\cite{moritzJATIS2024}

\begin{figure}
\begin{center}
\begin{tabular}{c}
\includegraphics[height=4.5cm]{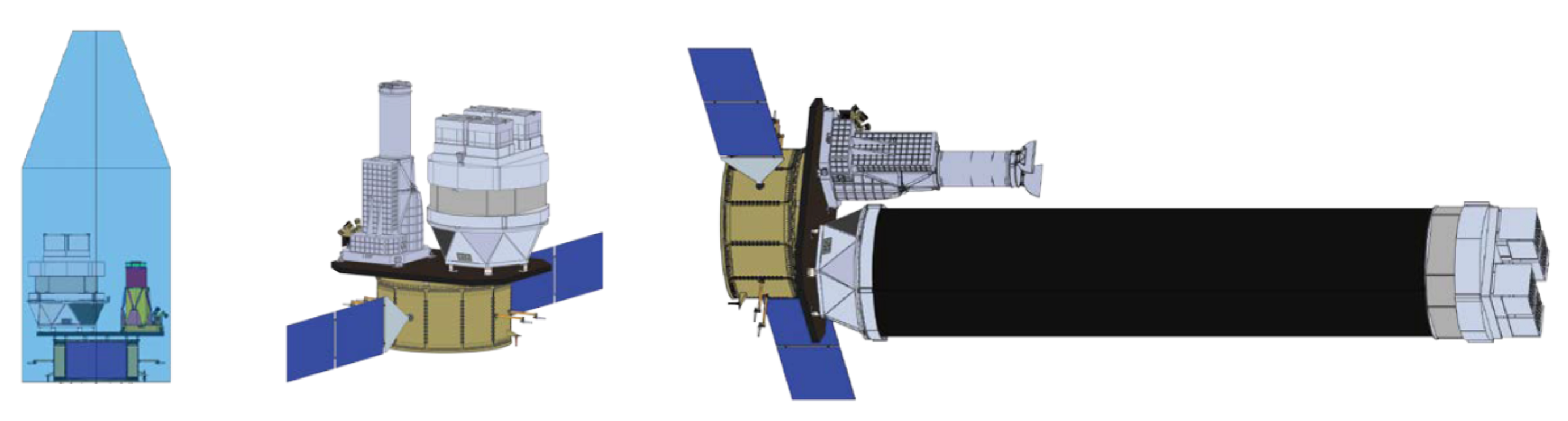}
\end{tabular}
\end{center}
\caption 
{ \label{fig:Probe}
Arcus Probe schematics.  Left: Spacecraft with instruments stowed in launch vehicle fairing. Middle: In space after solar panel deployment deployment.  The UV spectrometer is on the left, the XRS on the right.  Right: After XRS boom deployment.} 
\end{figure} 

The combination of optics and grating petal comprises an OC.  The four OCs are mounted to a common forward assembly (FA), which in turn is mounted to a coilable boom that is stowed during launch and deploys/uncoils after orbit insertion (see Fig.~\ref{fig:Probe}).  The boom is connected on the other end to the rear assembly (RA), which serves as the interface to the spacecraft and holds the two CCD detector subsystem assemblies (DSA), plus other systems.  Information about an earlier version of the boom can be found in Ref.~\citeonline{boom}.

\begin{figure}
\begin{center}
\begin{tabular}{c}
\includegraphics[height=4.0cm]{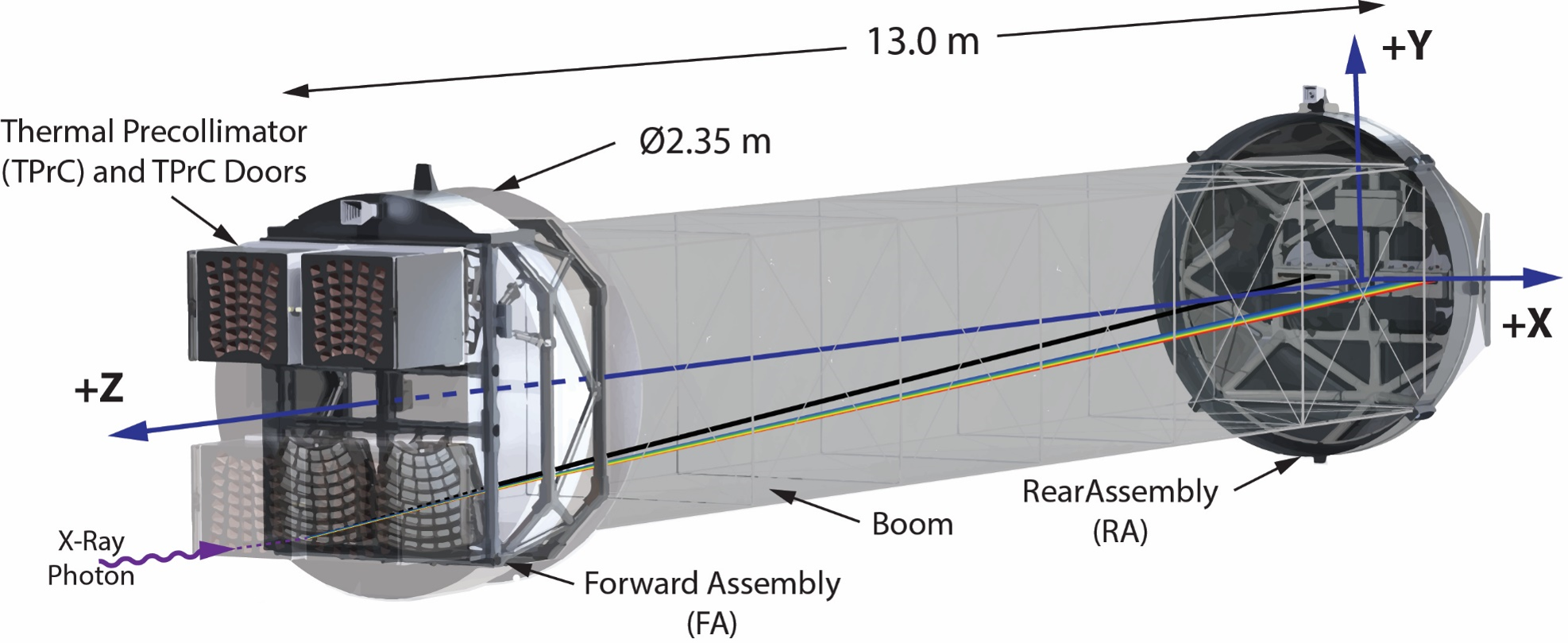}
\includegraphics[height=4.6cm]{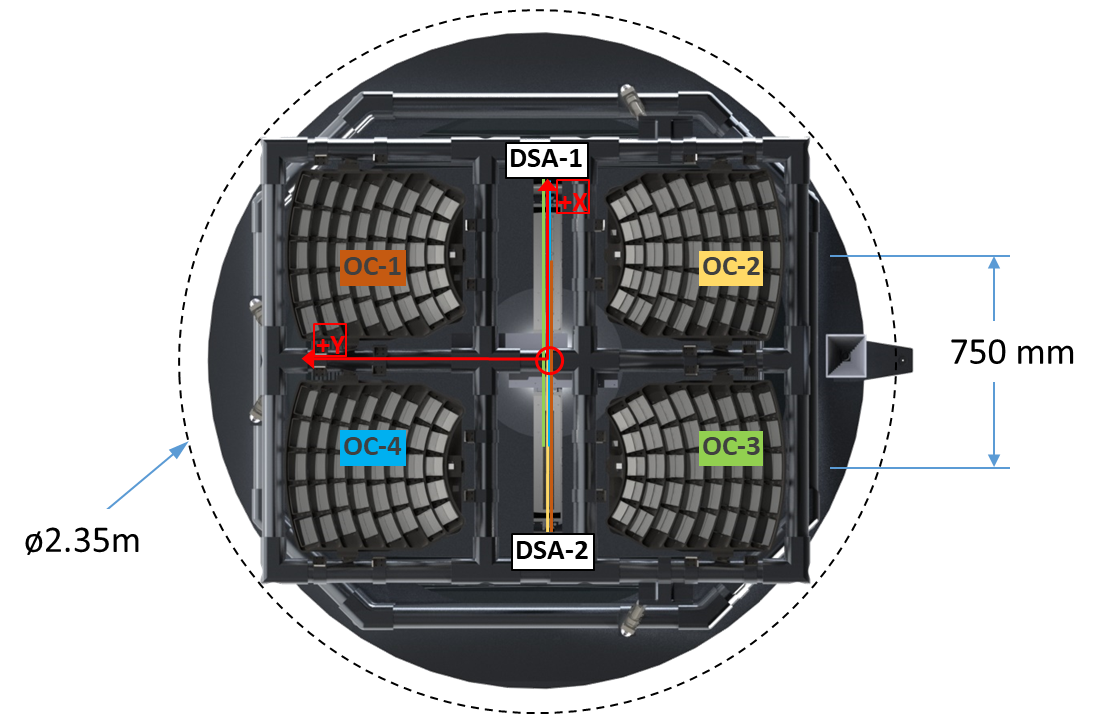}
\end{tabular}
\end{center}
\caption 
{ \label{fig:XRS}
Layout of the XRS. Left: Angled view, showing the thermal precollimators and SPO arrays for the four OCs on the left.  The grating petals are obscured behind the SPO arrays.  An x ray is shown that enters the lower left OC (OC-3) and either lands at the telescope focus (black line to DSA-2) or gets diffracted into a blazed order (rainbow colors to DSA-1). Right: View from the x-ray source into the XRS, showing the four OCs and the two DSAs. } 
\end{figure}

The gratings in each channel are placed along the surface of a tilted Rowland torus that also contains the telescope focus.  This guarantees that best focus in the dispersion direction for the non-zero orders is achieved, and also lies on the Rowland torus surface.  One DSA (DSA-1) sits at the focus of optical channels OC-1 and OC-2.  DSA-2 collects the strong blazed orders from OC-1 and OC-2.  Due to the symmetries of the design, DSA-2 can simultaneously lie on the differently tilted Rowland torus for OC-3 and OC-4 and their foci.  The gratings in the latter two channels are blazed in the opposite direction from the gratings in the first two channels, creating strong, focused diffraction orders on DSA-1 (see Fig.~\ref{fig:XRS}).  Each OC is offset from the others by a few mm in the cross-dispersion direction, leading to four quasi-(anti)-parallel spectra collected by only two compact DSAs.  This so-called double tilted Rowland torus design combines dense stacking of sub-apertured, narrow-LSF optics in a limited aperture, with a minimal number of readout CCDs.  It is described in detail in Refs. \citeonline{moritz2017} and \citeonline{moritzAPJ2024}.  The readout cameras are described in detail in a separate paper of this Special Issue.\cite{GrantJATIS2024} 

A grating facet consist of a $32.5 \times 32$ mm$^2$ silicon membrane aligned and bonded to a metal facet frame.\cite{SPIE2018}  All grating facets for all channels have the same design, are operationally identical and can be exchanged with each other.  Blazing is simply achieved through proper tilting relative to the incident x rays.  The grating facets are mounted to grating windows in groups of 4-6.  The windows are designed to cover the SPO beams with the minimum number of gratings of the given size.

\begin{figure}
\begin{center}
\begin{tabular}{c}
\includegraphics[height=5.0cm]{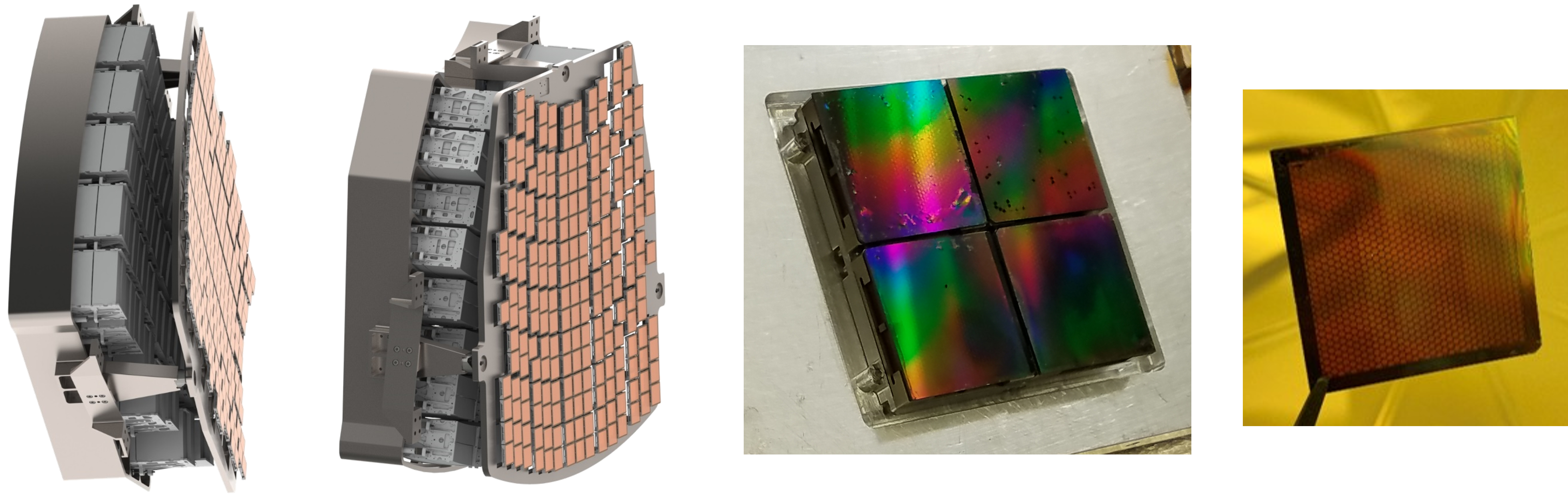}
\end{tabular}
\end{center}
\caption 
{ \label{fig:petalwindow}
Left: Two different views of an optical channel. The grating petal, holding 40 grating windows with a total of 216 grating facets, follows the shape of the Rowland torus.  Middle: Prototype grating window with four $32.5 \times 32$ mm$^2$ CAT grating facets.  Right: Silicon grating membrane, held by tweezers.} 
\end{figure} 

\section{Predicted XRS Performance}

The key performance parameters for the XRS are effective area ($A_{\mathrm {eff}}$) and resolving power ($R = \lambda/\Delta \lambda$), with $\lambda$ being the x-ray wavelength and $\Delta \lambda$ the smallest wavelength difference that can be resolved.  Effective area is mainly determined by the mirror effective area that feeds the grating array, grating diffraction efficiency (DE) summed over the collected diffraction orders, geometrical loss factors (x-ray blockage from support structures, gaps between gratings and CCDs), filter transmission, and CCD quantum efficiency.  Resolving power to first order is given by the diffraction angle of a given diffraction order, divided by the spectrometer LSF.  CAT gratings are blazed transmission gratings that have advantages in diffraction efficiency and resolving power in the soft x-ray band compared to existing grating instruments on Chandra\cite{cxc} and XMM/Newton.\cite{RGS}  In order to estimate XRS performance we need to understand the blazing principle underlying CAT grating design and the alignment tolerances that affect both $A_{\mathrm {eff}}$ and $R$.  Alignment tolerances are derived from detailed ray tracing that is described elsewhere in this Special Issue and for previous smaller versions of Arcus.\cite{moritzJATIS2024,moritz_SPIE2016,moritz2017,moritz_SPIE2018}

\section{CAT grating principle and structural hierarchy}

\begin{figure} [ht]
   \begin{center}
   \begin{tabular}{ c c c} 
   \includegraphics[height=6.5cm]{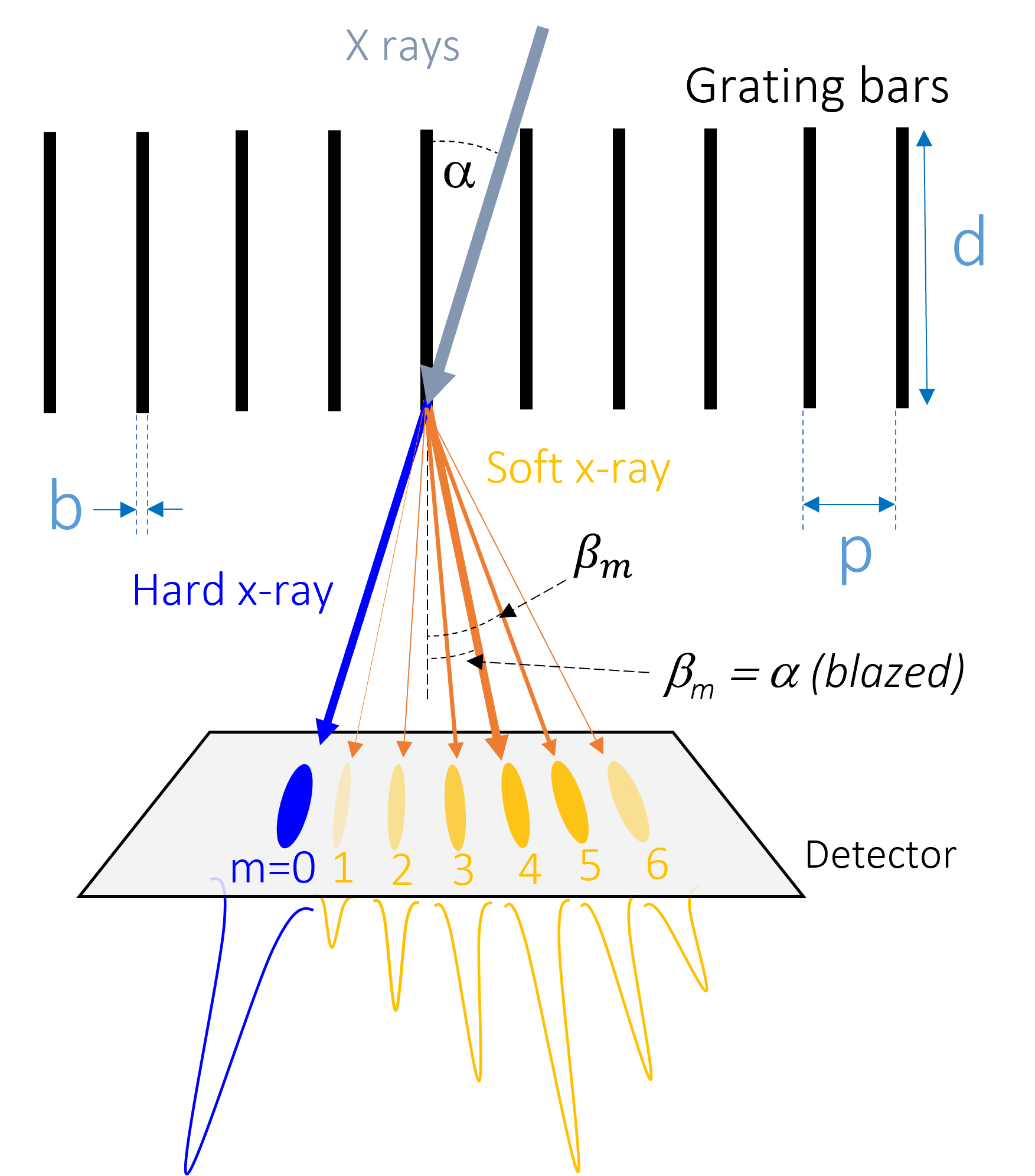}
   \includegraphics[height=2cm]{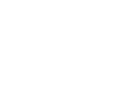}
   \includegraphics[height=5cm]{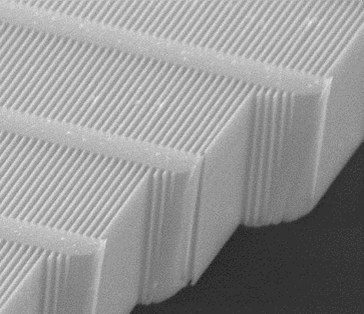}
   \end{tabular}
   \end{center}
   \caption
   { \label{fig:cross} 
Left: Schematic cross-section through a CAT grating of period $p$. The $m^{\rm {th}}$ diffraction order occurs at an angle $\beta_m$ where the path length difference between x rays from neighboring grating slots is $m\lambda$. Shown is the case where $\beta_m$ coincides with the direction of specular reflection from the grating bar sidewalls ($| \beta_m | = |\alpha|$), i.e., blazing in the $m^{\rm {th}}$ order. Right: Scanning electron micrograph (SEM) of a cleaved CAT grating membrane showing top, cross-section and sidewall views of the 200 nm-period silicon grating bars and their monolithically integrated 5 $\mu$m-period L1 cross supports (x rays enter from the top and leave out the bottom).   }
\end{figure}

CAT gratings feature ultra-high aspect-ratio, freestanding grating bars with nm-smooth sidewalls.\cite{Alex_JVST2013}  Each grating is inclined such that x rays of wavelength $\lambda$ impinge on the grating bar sidewalls at graze angles $\alpha$ below the critical angle for total external reflection $\alpha_c$ (see Fig.~\ref{fig:cross}).  The diffraction angle $\beta_m$ for the $m^{\mathrm {th}}$ diffraction order is given by the grating equation

\begin{equation}
{\frac{m \lambda}{p}} = \sin \alpha - \sin \beta_m ,
\label{ge}
\end{equation}

\noindent
where $p$ is the grating period.  Diffraction orders near the direction of specular reflection from the sidewalls show increased efficiency (i.e., blazing).  The small critical angles for soft x rays on the order of 1-2 degrees demand high-aspect ratio grating bars in order to intercept all incoming photons.  Furthermore, the bars should be as thin as possible to minimize blockage and absorption.  The grating period cannot be too large compared to the x-ray wavelength to obtain diffraction orders that can be sorted by order using the energy resolution of Si-based detectors.  The design initially fabricated for Arcus has a grating period $p = 200$ nm, grating bar depth $d = 4$ micrometers, and bar thickness $b \approx 60$ nm.  Recently we demonstrated $d > 5.5$ $\mu$m.\cite{SPIE2021}.  Blazing is most efficient when $\tan \alpha \approx (p-b)/d$. (For $d = 4$ $\mu$m this means $\alpha \approx 2$ deg.)

\begin{figure} [ht]
   \begin{center}
   \begin{tabular}{c  c  c } 
   \includegraphics[height=6cm]{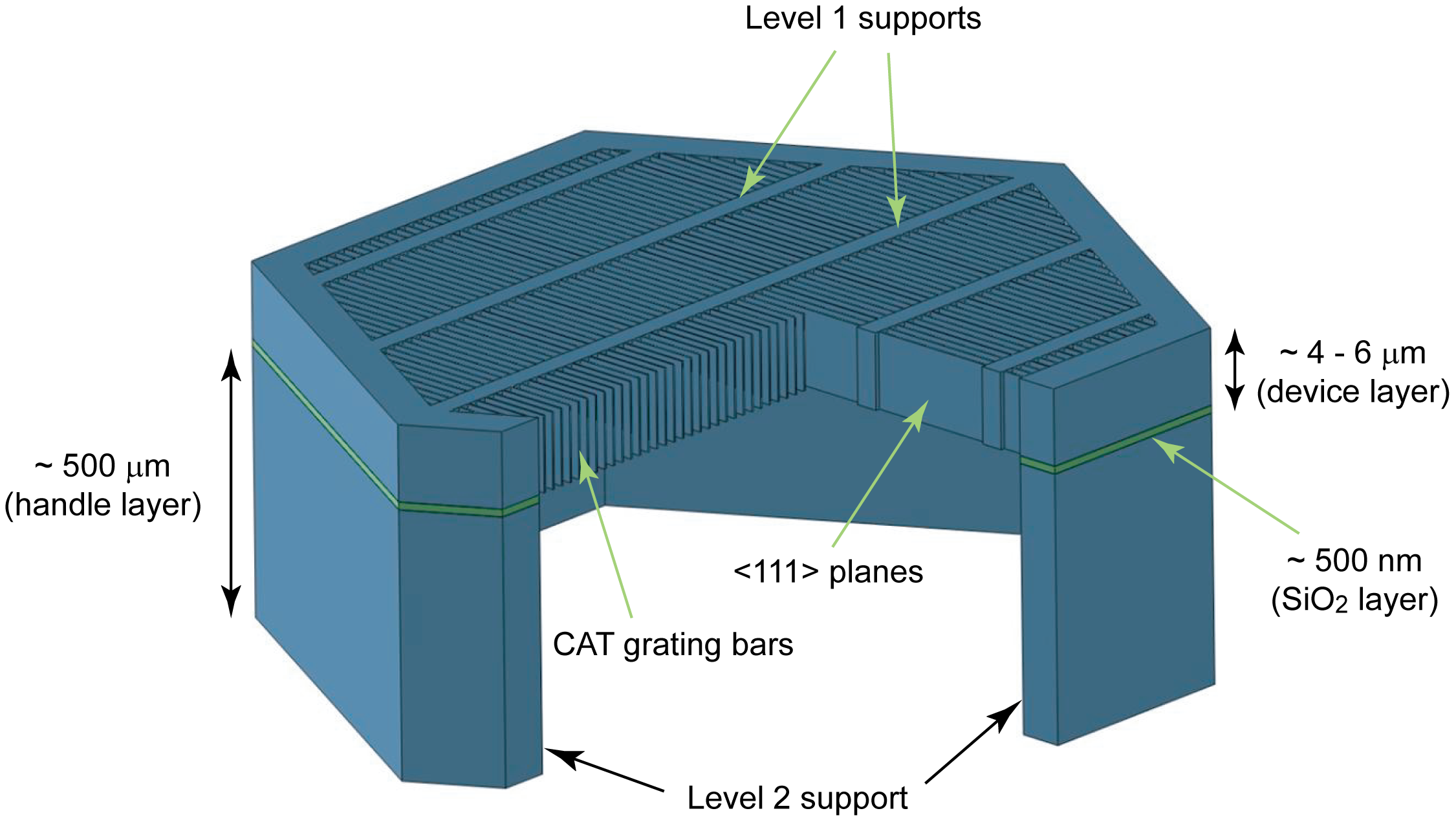}
   \end{tabular}
   \end{center}
   \caption
   { \label{fig:L1} 
Schematic showing the structural hierarchy of a CAT grating membrane (not to scale). See text for details.
}
\end{figure}

CAT grating bars are not supported by a membrane, but freestanding.  As seen on the right in Fig.~\ref{fig:cross}, the bars are held in place by a monolithically integrated 5 $\mu$m-period Level 1 (L1) support mesh.  Additional support structures are needed for the few-$\mu$m thin grating layer in order to manufacture large enough CAT gratings that can cover large areas on the order of thousands of square centimeters with a manageable number of gratings.  Fig.~\ref{fig:L1} shows the additional, much thicker and stronger Level 2 (L2) hexagonal support structure on the scale of $\sim 1$ mm.  

\section{CAT Grating Fabrication}

CAT grating fabrication has been described extensively in previous work and is briefly summarized here.\cite{Alex_SPIE2013,Alex_JVST2013,EIPBN2016,SPIE2020}
CAT gratings are currently made from 200 mm silicon-on-insulator (SOI) wafers.  The thin Si device (``front side") layer of the SOI wafer is manufactured to thickness $d$ (see Fig.~\ref{fig:cross}).  Using 193 nm 4X optical projection lithography (OPL) at MIT Lincoln Laboratory, patterns for CAT gratings, L1 and L2 structures are simultaneously transferred into a silicon oxide layer that serves as a mask for the subsequent deep reactive-ion etch (DRIE).  The $\sim 0.6$ mm thick handle (``back side") layer of the SOI wafer is then DRIE'd with a hexagonal pattern that is aligned with the front side L2 hexagons.  The grating bars are aligned parallel to the vertical \{111\} planes of the (110) device layer.  Since DRIE leaves rough sidewalls, this crystal orientation is used to ``polish" the grating bar sidewalls post-DRIE through immersion in KOH solution.\cite{Alex_JVST2013}  The gratings have to be dried in a critical-point dryer to prevent stiction due to liquid-vapor surface tension.  Finally, the buried oxide layer separating device and handle layers is removed in the areas not covered by Si, resulting in a monolithic Si grating layer with freestanding CAT grating bars, integrated L1 and L2 supports, and a bulky L2 mesh.

CAT gratings can be fabricated in volume manufacturing mode from 200 mm SOI wafers,\cite{SPIE2020} in principle allowing for $\sim 16-20$ gratings to be produced from a single wafer.

\section{CAT Grating X-ray Performance}

The main performance requirements for x-ray diffraction gratings are high diffraction efficiency and enabling high resolving power in a spectrometer instrument.  

\subsection{Diffraction Efficiency}

Diffraction efficiency has been measured repeatedly at beamline 6.3.2 of the Advanced Light Source at Lawrence Berkeley National Laboratory.  The focused, tunable monochromatic x-ray beam has small enough diameter to be placed within a single L2 hexagon, but it integrates over many tens of L1 periods.  A slit-covered photodiode detector is placed at the angle of a transmitted diffraction order, and the grating is rotated to measure DE over a range of several degrees in incidence angle.  The detector remains in place during grating rotation, since the change in diffraction angle as a function of grating rotation is negligible.  This is a major advantage of the transmission geometry, since this also means that certain grating alignment tolerances are very relaxed compared to reflection gratings.

\begin{figure} [ht]
   \begin{center}
   \begin{tabular}{c  c  c } 
   \includegraphics[height=4.8cm]{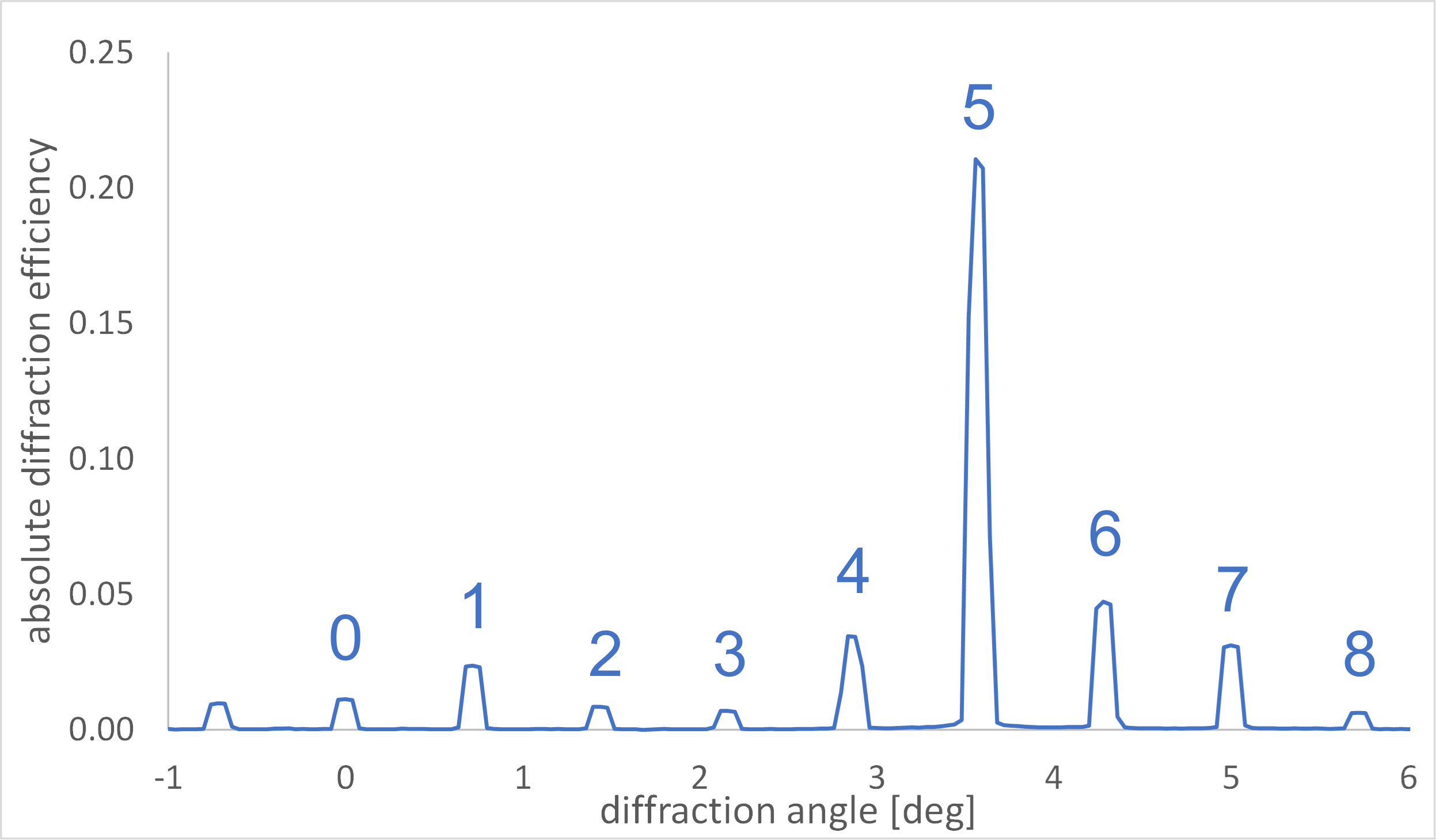}
   \includegraphics[height=4.8cm]{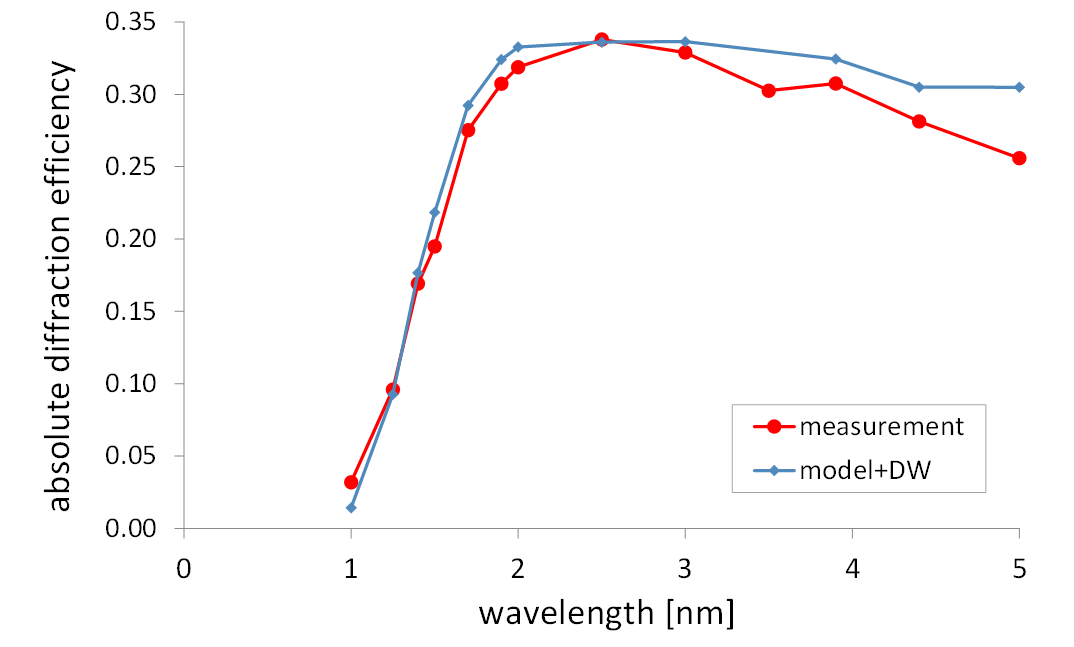} 
   \end{tabular}
   \end{center}
   \caption
   { \label{fig:eff} 
Representative diffraction efficiencies of 4 $\mu$m-deep CAT gratings from synchrotron data.  Left: Example detector scan for $\lambda = 2.5$ nm. DE is strongy blazed for $5^{\rm {th}}$ order at 21\%.  Summing over orders 4-7 gives a total DE of 32.2\%.  These values include absorption by L1 bars, but not blockage by the L2 mesh.  Right: Comparison of model efficiency (sum of blazed orders collected by XRS readouts), multiplied by Debye-Waller-like roughness factor, for 82\% L1 open area fraction with synchrotron spot measurements (sum of same blazed orders as for model) for a $32 \times 32$  mm$^2$ CAT grating\cite{SPIE2017}.  The L2 mesh is not considered in this plot.
}
\end{figure}

The earliest x-ray testable freestanding CAT grating structures had larger period ($p = 574$ nm) and bulky L1 cross supports that were $>> 1$ $\mu$m wide.\cite{OE2008,SPIE2008}  DE can be modeled using rigorous coupled-wave analysis (RCWA).\cite{RCWA}   Typically, CAT gratings perform in the range of 80-100\% of RCWA predictions for DE, where the difference from 100\% can often be modeled with a Debye-Waller-type (DW) roughness factor.\cite{SPIE2020}.  Current prototypes with 200 nm period, $d = 4$ $\mu$m, and $b \approx 60-70$ nm deliver DE $> 20$\% in individual orders and $> 30$\% when summed over the strongest (``blazed") orders ($\beta_m \sim 2\alpha$) near O-K wavelengths (see left of Fig.~\ref{fig:eff}).  These numbers include losses caused by absorption from the L1 supports.

Several 4 $\mu$m-deep gratings were measured at several wavelengths, and the data is modeled with RCWA and a DW factor (see right of Fig.~\ref{fig:eff}).  The model is then used to extrapolate efficiency between the measured wavelengths and angles as an input for ray-trace-based effective area predictions for the XRS.  Ray tracing also takes into account alignment tolerances and the changing efficiencies as a function of grating angles.\cite{moritz_SPIE2023,moritzJATIS2024}

The synchrotron beam can be scanned over a whole grating to examine uniformity of efficiency.\cite{SPIE2015,SPIE2016,SPIE2017,SPIE2018}  Alternatively, we also measured efficiency in the converging beam of an SPO, almost fully illuminating a large CAT grating in an Arcus-like configuration, and verified that it agreed with synchrotron spot measurements and exceeded model assumptions.\cite{ApJ_2022}

\subsection{Resolving Power}

Resolving power $R$ is an expression of how much a spectral line is broadened by the XRS response function.  The main ingredients are the mirror LSF (PSF projected onto the dispersion axis), the diffraction angles of the measured orders, the distance of the gratings from focus, the aberrations of the optical design, the CCD pixel size, and grating imperfections.  Large $R$ requires large grating distance from focus and large diffraction angles, and small values for the other terms.  

The most obvious grating imperfection that would limit $R$ is a variation $\Delta p$ of grating period $p$, since it would introduce a variation in $\beta_m$ (see Eq.~\ref{ge}).  For a Gaussian distribution of $\Delta p$ one can define an effective grating resolving power $R_g = p/\Delta p$ as an additional term in the instrument response function that contributes to the broadening of measured spectral lines.\cite{ApJ_2022}  $R_g$ is an upper limit to the resolving power of the XRS and can be measured in the following way:  First, the PSF of a focusing optic (the ``direct beam") is measured with a narrow-line soft x-ray source, such as the well-characterized Al-K$_{\alpha}$ doublet.  Then, a grating is inserted into the beam, and the source spectrum is measured in the highest accessible diffraction order.  The measured spectrum is a convolution of the known source spectrum with the measured LSF of the direct beam and a Gaussian of width $\Delta \beta_m$ caused by $\Delta p$.  Thus, 
$\Delta p$ and $R$ can be extracted from fitting to the measured spectrum.  This measurement has been performed up to 18$^{\rm {th}}$ order at the Marshall Space Flight Center Stray Light Facility using a slumped glass mirror pair\cite{SPIE2016,SPIE2017,AO2019} and up to 21$^{\rm {st}}$ order at the PANTER x-ray facility using an SPO.\cite{SPIE2020,ApJ_2022}  In all cases $R_g \approx 10^4$ was found, which significantly exceeds the $R_g > 3850$ Arcus requirement. 

\begin{figure} [ht]
   \begin{center}
   \begin{tabular}{ c c c} 
   \includegraphics[height=7.0cm]{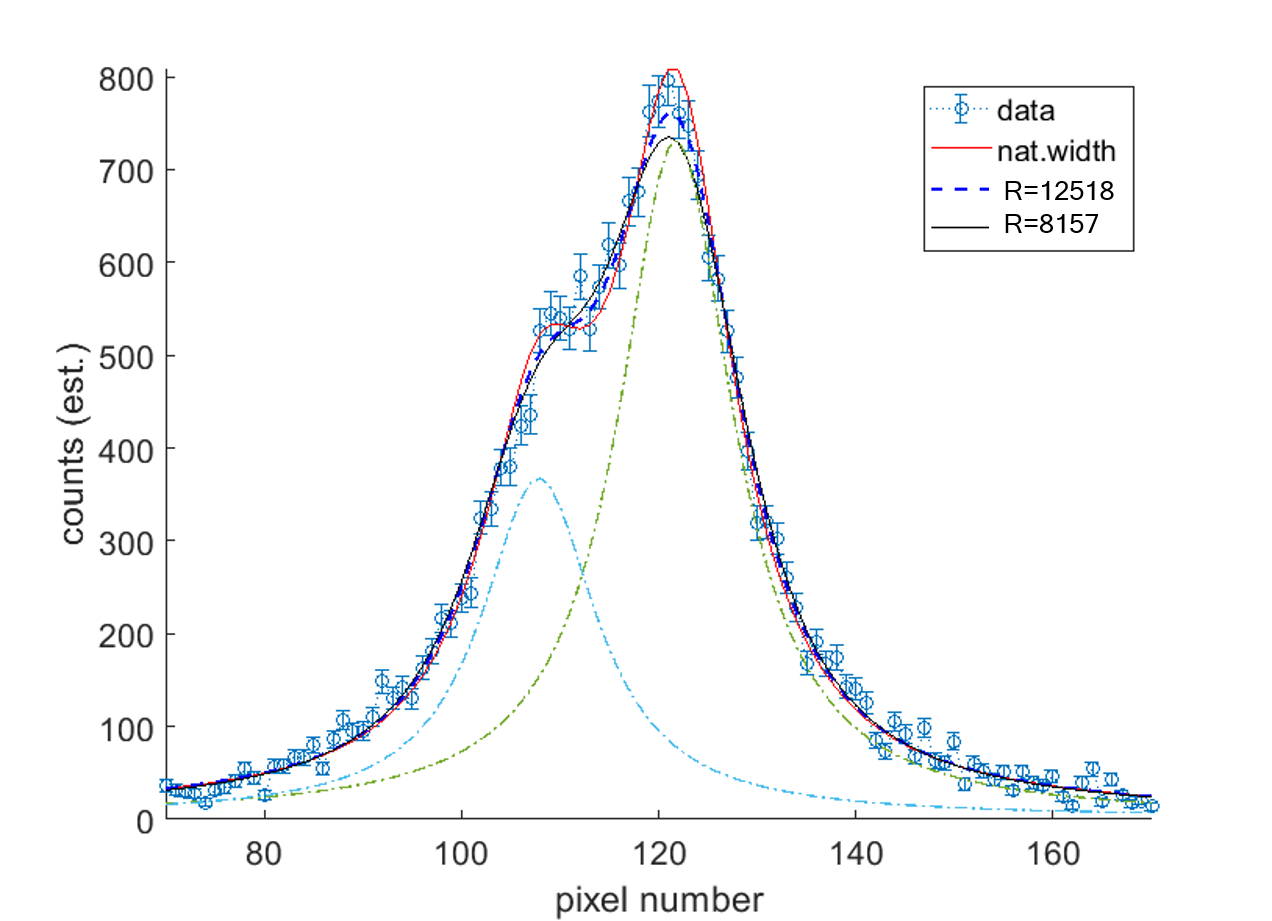}
   \end{tabular}
   \end{center}
   \caption
   { \label{fig:R} 
 Spectrum of Al anode from two mutually aligned and simultaneously illuminated CAT gratings, showing the Al-K$_{\alpha}$ doublet measured in $18^{\mathrm {th}}$ order.\cite{ApJ_2022}  Camera pixels are 20 $\mu$m in size.  The red line is the natural line shape of the doublet,\cite{AO2019} and the dashed line is the best fit to the data. The black solid line is the curve for the lower $3\sigma $ confidence limit, corresponding to $R_G = 8157$. The upper $3\sigma $ confidence limit includes $R = \infty$.
 The dash-dotted lines show the individual K$_{\alpha_1}$ and K$_{\alpha_2}$ components with their natural widths on top of the weak sloped background. }
\end{figure} 

A linear array of four co-aligned CAT gratings, illuminated by a pair of confocal SPOs in an Arcus-like setup, also performed as expected.\cite{SPIE2018}.  As shown in Fig.~\ref{fig:R}, we recently derived $R_g \approx 1.3^{+\infty}_{-0.5} \times 10^4$ ($3\sigma$) for a different pair of co-aligned gratings, simultaneously illuminated by a single SPO with LSF $\sim 1$ arcsec (full width half max - FWHM).\cite{ApJ_2022}

The limiting factor in resolving power for the Arcus XRS therefore is not the quality of CAT gratings, but the combined LSF from the whole OC SPO petal, required to be $< 3.4$ arcsec (FWHM), and expected to be 2.6 arcsec (FWHM).

\subsection{XRS Figures of Merit}

It is useful to discuss the leap in performance that CAT grating technology provides in the soft x-ray band over previous x-ray grating technologies.

\begin{figure} [ht]
   \begin{center}
   \begin{tabular}{ c c c} 
   \includegraphics[height=5.35cm]{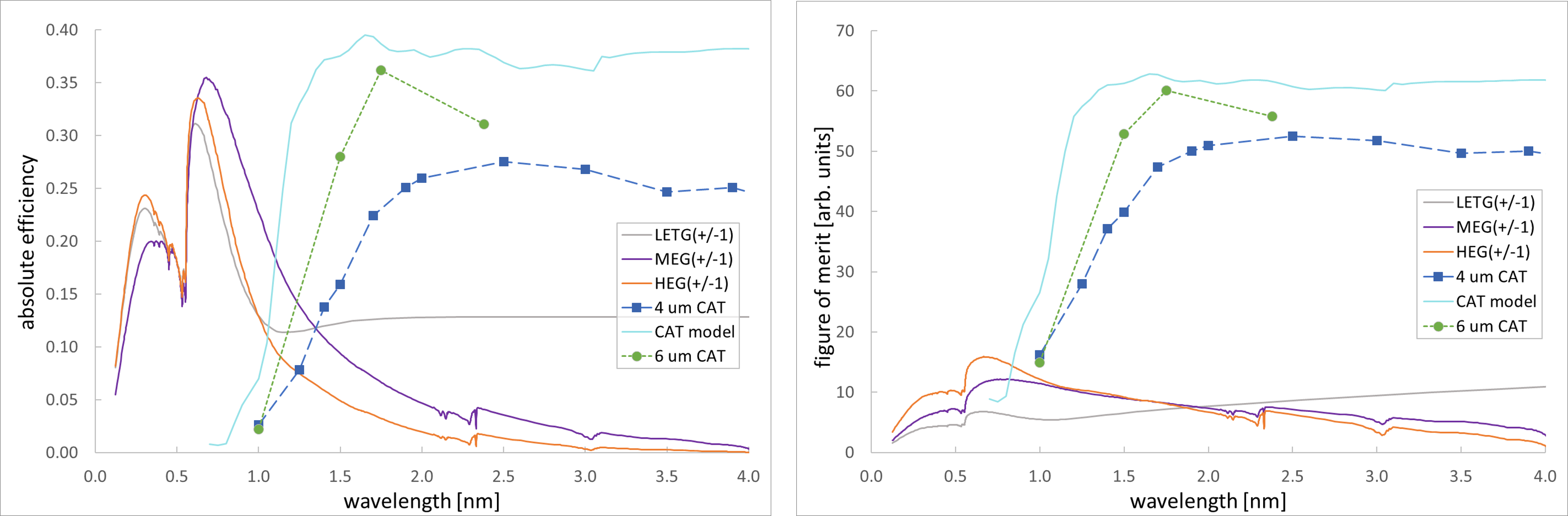}
   \end{tabular}
   \end{center}
   \caption
   { \label{fig:TE_DE} 
Comparison of TG diffraction efficiencies and figure of merit.  {\bf Left:} For Chandra gratings the sum of $\pm 1^{\mathrm {st}}$ order DEs from calibration files is shown \cite{marshall:2012}.  These DEs include all losses from support structures and gaps between facets.  For CAT gratings all orders within the blaze envelope are summed up, and a total loss of 31\% from support structures and gaps is assumed. ``4 $\mu$m CAT" and ``6 $\mu$m CAT" are from synchrotron measurements of CAT gratings with $d = 4$ and $\sim 6$ $\mu$m.  The latter are from the first prototypes of this thickness.\cite{SPIE2021,SPIE2023}  ``CAT model" is the sum of blazed order DEs from a model with $d = 6$ $\mu$m, $b = 40$ nm, and 1.5 nm DW roughness parameter.
 {\bf Right:} Figure of merit for weak line detection 
 ($\sim \sqrt{A_\mathrm{eff} R}$).  The same mirror effective area and constant detector efficiency is assumed for all plots.  For CAT gratings, $R = 10^4$ is assumed as an upper limit, based on conservative analysis of lab measurements \cite{SPIE2016,AO2019,ApJ_2022}. The HETG and LETG instruments complement each other over the shown range.  Chandra with CAT gratings would be superior by far for $\lambda > 1$ nm due to a combination of higher DE and blazing into higher orders. }
\end{figure}

The High and Low Energy Transmission Grating Spectrometers (HETGS\cite{cxc}, $p = 200$ (HEG) and 400 nm (MEG); and LETGS, \cite{Predehl_1992} $p = 991$ nm) on Chandra have the advantage of small mirror LSF on the order of 0.5 arcsec, but most of the diffracted photons land at small angles in $m = \pm 1^{\rm {st}}$ orders ($\beta_m \approx \lambda/p = 1$-$25 \times 10^{-3}$ for the 1-5 nm wavelength band).  For the Arcus XRS, CAT gratings are inclined by 1.8 degrees relative to the incident x rays, and the strongest orders are found near $\beta_m = 0.063$, relatively independent of $\lambda$.  Chandra gratings were designed as phase-shifting gratings with high efficiencies below 1 nm, and peaking near 0.7 nm wavelength.  Fig.~\ref{fig:TE_DE} compares diffraction efficiencies for different grating types.  CAT gratings are designed for broadband high efficiency for wavelengths longer than 1 nm.  Arcus assumes performance shown as ``4 um CAT".  The ``6 um CAT" curve shows results from deeper, more recent gratings with $d \sim 5.5$-6 $\mu$m that could be used instead, which would improve Arcus performance further.  The right side of Fig.~\ref{fig:TE_DE} is an ``apples-to-apples" comparison, assuming CAT gratings were placed on Chandra.  It shows a figure of merit for the detection of absorption lines, which is proportional to $\sqrt{A_{\rm {eff}} \times R}$.  We conservatively limited $R$ to $10^4$, even though the Chandra mirror LSF would allow for higher values of $R$.  CAT gratings outperform Chandra gratings significantly for $\lambda > 1$ nm.

\begin{figure} [ht]
   \begin{center}
   \begin{tabular}{ c c c} 
   \includegraphics[height=5.8cm]{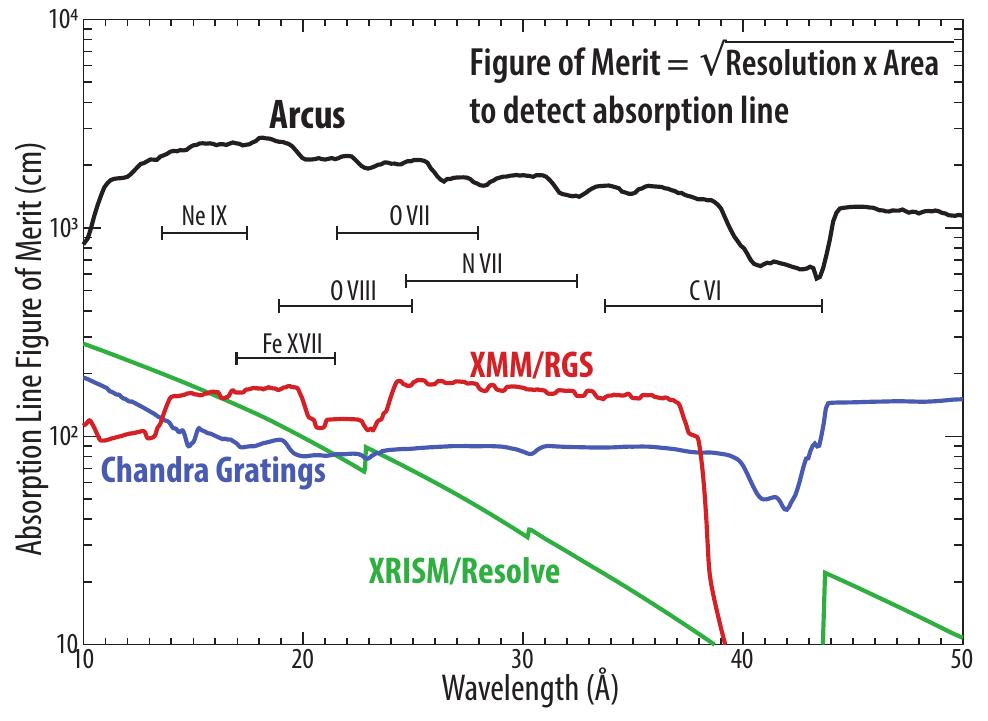}
   \includegraphics[height=5.8cm]{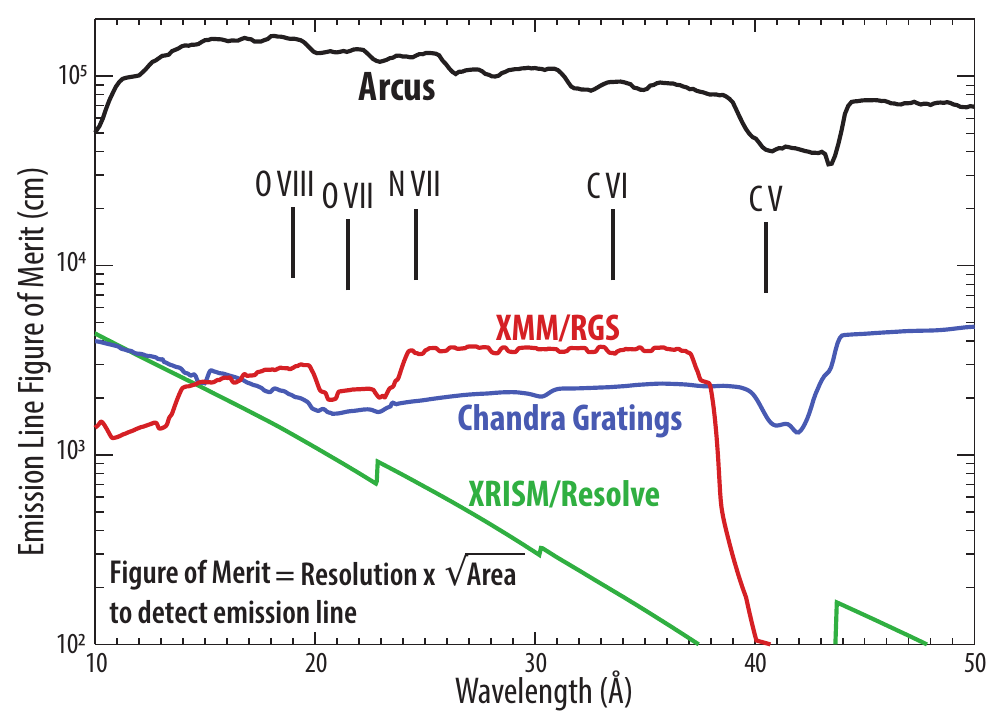}
   \end{tabular}
   \end{center}
   \caption
   { \label{fig:FOM} 
Comparison of figures of merit between soft x-ray spectroscopy instruments in operation and the Arcus Probe XRS.  ``Chandra Gratings" refers to the better of HEG, MEG, and LETG.  The curve for the Resolve microcalorimeter on XRISM\cite{Resolve2022} assumes that the closed aperture door is open.  The wavelength ranges for important plasma diagnostics lines are also indicated.  Left:  Figure of merit for absorption line detection.  Right: Figure of merit for emission line detection.  }
\end{figure} 

Fig.~\ref{fig:FOM} compares two different figures-of-merit for actual instruments on different missions with the predicted Arcus performance.  Despite the wider LSF compared to Chandra, the Arcus XRS is expected to exceed Chandra HETG and XMM/Newton RGS figures of merit by 1-2 orders of magnitude across the 1-5 nm band, due to a combination of CAT grating properties and the larger mirror effective area.  Also shown is the Resolve instrument on XRISM, which is a microcalorimeter with fixed energy resolution of slightly below 5 eV. It provides better performance at much shorter wavelengths than shown.\cite{Resolve2022}

\section{CAT Grating Alignment}

Alignment tolerances for CAT gratings are derived using ray-tracing models as described in Refs.~\citeonline{moritzJATIS2024,moritz_SPIE2016,moritz2017,moritz_SPIE2018}.  Positioning tolerances are in the 200 $\mu$m to mm ($3 \sigma$) range, well within precision machining capabilities.  Two rotational degrees of freedom require custom metrology for alignment before the grating membrane is bonded to its metal facet frame:

Grating roll (rotation around the grating normal, which also rotates the dispersion axis) can be measured using visible light diffraction from the L1 cross support mesh,\cite{SPIE2021} which is defined in the OPL mask to be oriented at 90 degrees from the CAT bars.\cite{SPIE2020}  Before bonding, roll relative to the facet frame sides is adjusted to be the same for all grating facets, using a reference grating facet.  

Grating yaw, which controls the angle of incidence onto the CAT grating bar sidewalls, requires a separate measurement of the grating bar tilt relative to the grating membrane surface normal using small-angle x-ray scattering (SAXS).\cite{Song_SPIE2018,JungkiEIPBN}  Using laser reflection, the measured bar tilt is then compensated for during the bonding of the Si membrane to its metal frame, such that the average grating bar direction ends up being parallel to the normal of the facet frame bottom, which is the mechanical reference surface for mounting to the grating window.  The success of this alignment method within Arcus tolerances was demonstrated with x rays for a $2\times2$ grating window.\cite{SPIE2022,ApJ_2022}

\section{Increasing Diffraction Efficiency through Bar Thinning}

Besides fabricating CAT gratings with $d > 4$ $\mu$m and reducing the cross sections of support structures, XRS effective area can also be improved by reducing the grating bar width $b$.  The fabrication method described above typically results in bar widths in the 60 nm range.  Smaller widths could in principle be achieved by creating an oxide mask with lower duty cycle (the ratio of mask line width to grating period).  However, the resulting thinner bars can suffer from increased damage and destruction during the more aggressive wet etching and cleaning steps.  Obviously, thicker bars are less fragile and preferred during wet processing steps.

Grating bar width can be reduced after the above fabrication steps using repeated cycles of oxidation - which consumes a small amount of Si - and oxide removal using HF vapor.  Both steps are gas-based and gentle.  In a first set of experiments we demonstrated bar thinning on gratings that were not freestanding, simply by removing native oxide with HF vapor, letting the native oxide layer reform in air, and repeating the cycle.\cite{SPIE2022}  We subsequently repeated the experiment on three freestanding gratings, after measuring their DE.  Thinning is observable in top and bottom SEM images after about 10 cycles.  Here we present first comparisons of DE before and after bar thinning.

\begin{figure} [ht]
   \begin{center}
   \begin{tabular}{ c c c} 
   \includegraphics[height=5.8cm]{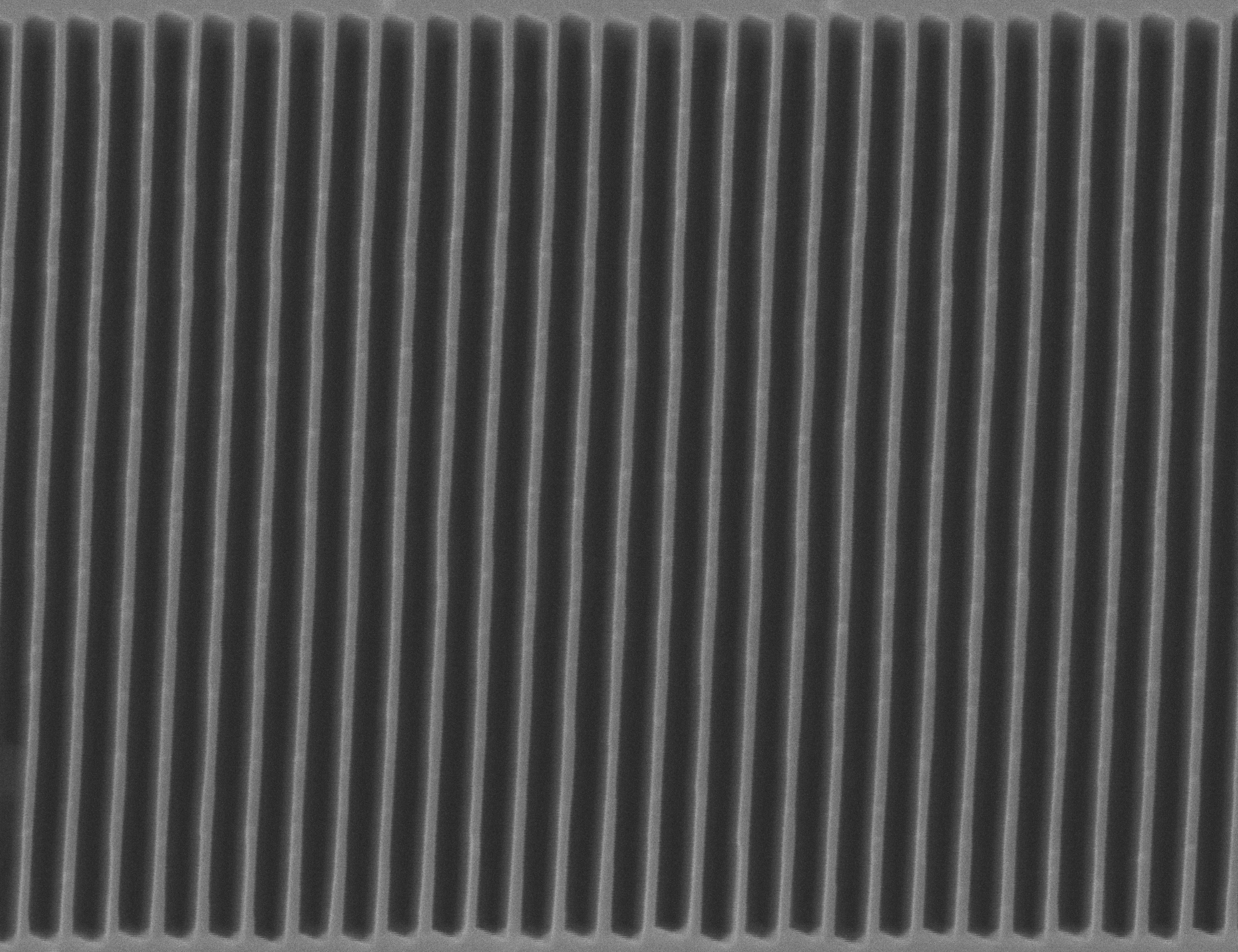}
   \includegraphics[height=5.8cm]{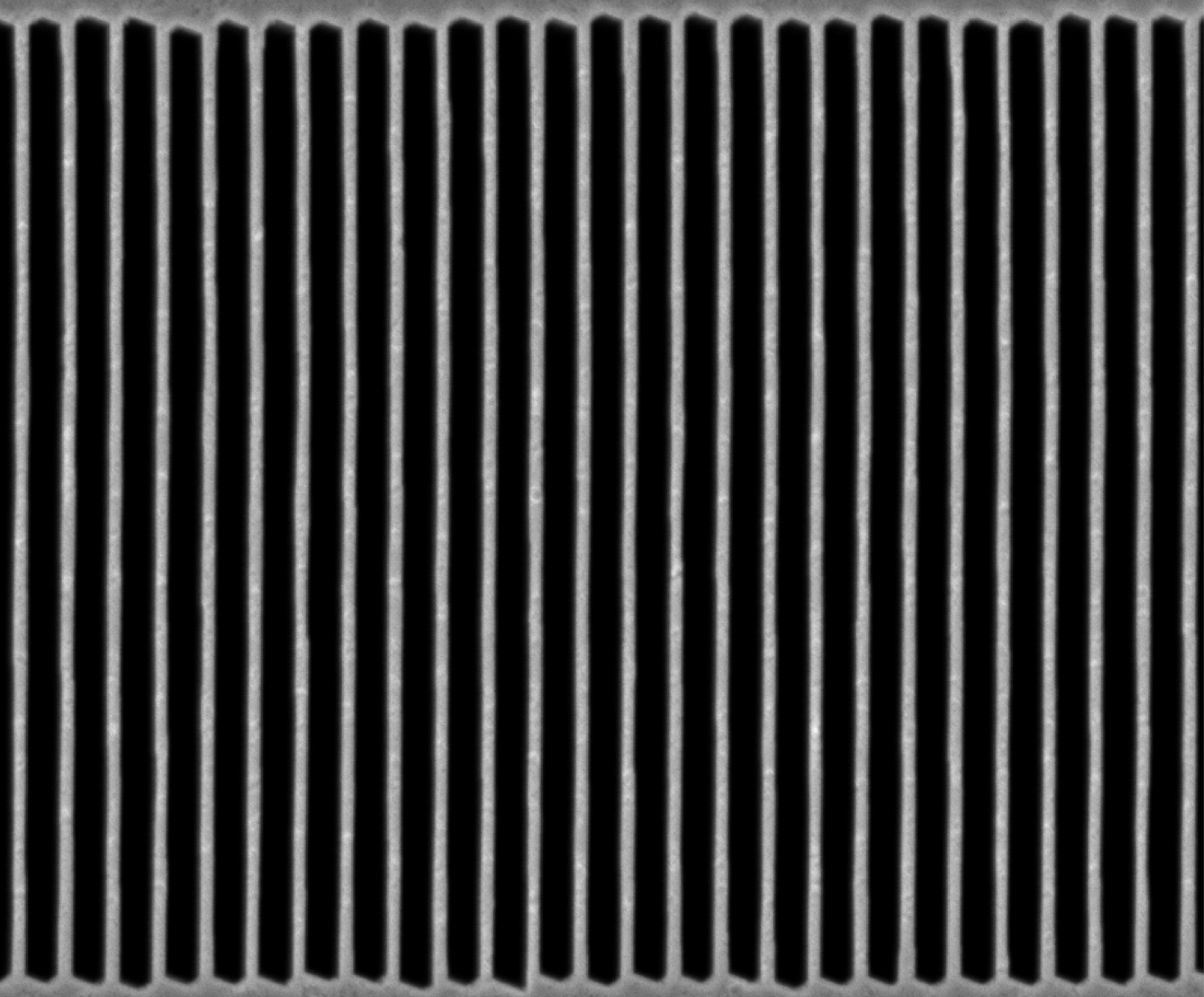}
   \end{tabular}
   \end{center}
   \caption
   { \label{fig:thin} 
   Top down scanning electron micrograph (SEM) images of grating SP3 before (left) and after 20 cycles of HF vapor etching and oxidation in ambient air (right).  The 200 nm-period grating bars are slightly thinner after 20 cycles, but the difference is difficult to quantify precisely from SEM images alone.
 }
\end{figure}

Of the three tested gratings with $d \sim 6$ $\mu$m, one each underwent 10 (SP1), 20 (SP3) and 30 (SP5) cycles of HF vapor oxide removal and native oxide regrowth.  The oxide was allowed to reform for at least 24 hours before the next HF vapor etch.  Fig.~\ref{fig:thin} shows top-down SEM images of grating SP3 before and after treatment.  After image analysis we estimate the CAT grating bars to be $\sim 4 \pm 2$ nm thinner, changing from $\sim 57$ to $\sim 53$ nm at the top.  Changes in thickness $b$ deeper into the gratings obviously can not be discerned from SEM images.  For grating SP1 we can not detect any clear changes.  The changes for grating SP5 look similar to the ones for SP3.

\begin{figure} [ht]
   \begin{center}
   \begin{tabular}{ c c c} 
   \includegraphics[height=5.8cm]{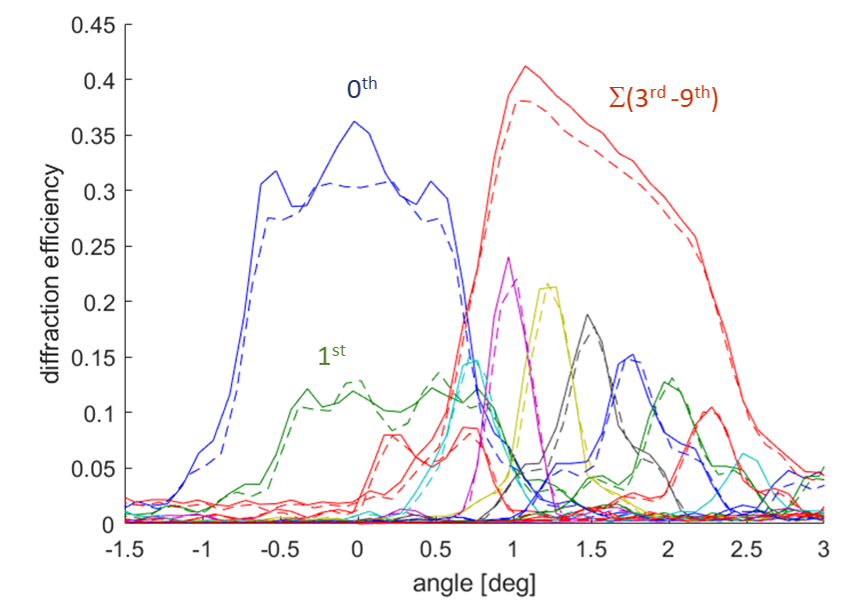}
   \includegraphics[height=5.8cm]{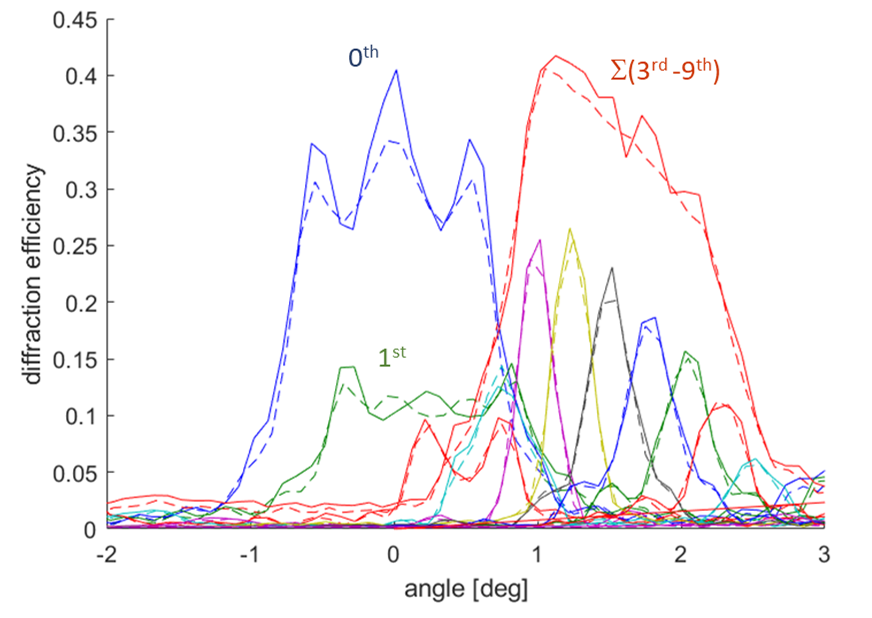}
   \end{tabular}
   \end{center}
   \caption
   { \label{fig:xthin} 
   Diffraction efficiencies at $\lambda = 1.75$ nm for orders 0-10 and sum of orders 3-9 as a function of incidence angle.  Dashed lines are before thinning, solid lines after thinning.  Left: Grating SP3.  Right: Grating SP5.
 }
\end{figure}

The gratings were measured for diffraction efficiency at the synchrotron eleven months after their initial synchrotron measurements, and about two months after the start of the HF vapor etching cycles.  Measurements were taken inside the same L3 hexagon as before on each sample, except for SP1, which unfortunately was damaged during handling in the area of the previously measured hexagon. For SP1 another hexagon near the original location was selected, and we could not discern a systematic increase or decrease of DE. For gratings SP3 and SP5 we show DE comparisons in Fig.~\ref{fig:xthin}.  One can clearly see an increase in $0^{\rm {th}}$ order efficiency, especially at normal incidence, which indicates less blockage by Si.  For individual higher diffraction orders the DE increase is less pronounced, but it is clearly visible when summing over the blazed orders.  For both gratings we find an increase in blazed efficiency in the range of 2-3\% in the angular range where blazing is most efficienct.  Preliminary modeling of the DE with RCWA indicates that the CAT grating bars became $\sim $ 4-5 nm thinner on average, with average $b$ approaching $\sim$ 42-44 nm.  The data also show that DE is significantly higher than assumed for Arcus (see Fig.~\ref{fig:eff}).

While it is clear that HF vapor etching of silicon oxide leads to thinner grating bars, the etch could also change the roughness of the sidewalls, thereby increasing (lower roughness) or decreasing (higher roughness) DE.    It is unclear why 10 treatment cycles seem to have caused little change, and why we see little difference in outcome between 20 and 30 etch/oxidation cycles.  We plan to perform more systematic experiments, including fast growth of thicker oxide layers at elevated temperatures, to understand the different trade-offs between experimental conditions and outcomes.

There will be a mechanical limit to making grating bars thinner, when the grating membrane could get damaged by launch vibrations.  Vibration and temperature cycling testing so far have not revealed any problems,\cite{SPIE2017} but will have to be repeated for thinner and thinner structures.

\section{Future Work}

Future improvements in DE are possible through thinner CAT grating bars and thinner L1 and L2 support structures.  The Arcus Probe XRS assumes an L1 duty cycle of 18\%, but large CAT gratings with 10\% duty cycle have been fabricated previously\cite{SPIE2019} and could be used for Arcus.  The L2 open area fraction could be increased from the current 81\% to a higher value.  Deeper gratings also show higher DE, but the difference with 4 $\mu$m deep gratings is most pronounced at smaller incidence angles, which are preferred for larger mission concepts with smaller optics PSF, such as Lynx.\cite{Gaskin_JATIS2019,moritz_JATIS2019}  All of these improvements are actively being investigated.

\section{Summary}

Arcus Probe is a mission concept that features two high-resolution spectrometers.  The XRS described in this work is an instrument with 1-2 orders of magnitude improved performance compared to existing missions in the 1-5 nm soft x-ray band.  The four parallel OCs provide large effective area and high redundancy in a small footprint.  Alignment tolerances are well-understood and well within metrology and manufacturing capabilities.  Co-alignment with the UV spectrometer (UVS) is described in another paper in this Special Issue,\cite{CheimetsJATIS2024} as well as the UVS itself.\cite{FranceJATIS2024}

SPOs and CAT gratings already perform at required levels or exceed them.  A volume manufacturing approach utilizing tools from the semiconductor and MEMS industries has been developed that can produce the required 864 CAT gratings efficiently in two years.  Mechanical structures (facet frames, grating windows, grating petals, etc.) are standard precision engineering items.

\subsection*{Disclosures}

Several authors of this paper are members of the Arcus collaboration;  Should NASA select Arcus for implementation, their institutions will receive funding which may be used to fund the authors' salaries in full or in part in the future.

\subsection* {Code, Data, and Materials Availability} 
Data presented in this paper can be made available by the author (RKH) upon reasonable request.

\subsection* {Acknowledgments}
This work has been funded in part under NASA grants 80NSSC22K1904 and 80NSSC250K0780 and by the MIT Kavli Insititute for Astrophysics and Space Research.
We appreciate facility support from MIT.nano.  This research also used resources of the Advanced Light Source (beamline 6.3.2), a U.S. DOE Office of Science User Facility under contract no.~DE-AC02-05CH11231.


\bibliography{CAT_references}   

\begin{thebibliography}{10}

\bibitem{Astro2020}
{National Academies of Sciences, Engineering, and Medicine}, ``Pathways to {D}iscovery in {A}stronomy and {A}strophysics for the 2020s,''  (2021).

\bibitem{Arcus2023}
R.~K. Smith, ``{The {A}rcus {P}robe mission},'' in {\em UV, X-Ray, and Gamma-Ray Space Instrumentation for Astronomy XXIII},   {\bf 12678}, International Society for Optics and Photonics, SPIE  (2023).

\bibitem{SPO_SPIE2023}
B.~Landgraf, L.~Abalo, N.~M. Barri{\`e}re, {\em et~al.}, ``{High-resolution and light-weight silicon pore x-ray optics},'' in {\em Optics for EUV, X-Ray, and Gamma-Ray Astronomy XI},  S.~L. O'Dell, J.~A. Gaskin, G.~Pareschi, {\em et~al.}, Eds.,  {\bf 12679}, 1267903, International Society for Optics and Photonics, SPIE  (2023).

\bibitem{NewAthenaSPO-SPIE2023}
M.~Bavdaz, E.~Wille, M.~Ayre, {\em et~al.}, ``{NewATHENA optics technology},'' in {\em Optics for EUV, X-Ray, and Gamma-Ray Astronomy XI},  S.~L. O'Dell, J.~A. Gaskin, G.~Pareschi, {\em et~al.}, Eds.,  {\bf 12679}, 1267902, International Society for Optics and Photonics, SPIE  (2023).

\bibitem{Cash1991}
W.~Cash, ``X-ray optics .2. a technique for high-resolution spectroscopy,'' {\em APPLIED OPTICS} {\bf 30}, 1749--1759  (1991).

\bibitem{SPIE2010}
R.~K. Heilmann, J.~E. Davis, D.~Dewey, {\em et~al.}, ``Critical-angle transmission grating spectrometer for high-resolution soft x-ray spectroscopy on the {International X-ray Observatory},'' in {\em SPACE TELESCOPES AND INSTRUMENTATION 2010: ULTRAVIOLET TO GAMMA RAY},  M.~Arnaud, S.~Murray, and T.~Takahashi, Eds., {\em Proceedings of SPIE} {\bf 7732}, SPIE  (2010).
\newblock Conference on Space Telescopes and Instrumentation 2010 - Ultraviolet to Gamma Ray, San Diego, CA, JUN 28-JUL 02, 2010.

\bibitem{GrantJATIS2024}
C.~Grant, ``Arcus focal plane cameras.'' submitted to JOURNAL OF ASTRONOMICAL TELESCOPES INSTRUMENTS AND SYSTEMS  (2024).

\bibitem{moritzJATIS2024}
H.~M. G\"unther and R.~K. Heilmann, ``Arcus x-ray telescope performance predictions and alignment requirements.'' submitted to JOURNAL OF ASTRONOMICAL TELESCOPES INSTRUMENTS AND SYSTEMS  (2024).

\bibitem{boom}
H.~Bergner, P.~Cheimets, E.~Hertz, {\em et~al.}, ``{Development of a 12m coilable boom for the Arcus MIDEX mission.},'' in {\em UV, X-Ray, and Gamma-Ray Space Instrumentation for Astronomy XXII},  O.~H. Siegmund, Ed.,  {\bf 11821}, 118211E, International Society for Optics and Photonics, SPIE  (2021).

\bibitem{moritz2017}
H.~M. G\"unther, P.~N. Cheimets, R.~K. Heilmann, {\em et~al.}, ``Performance of a double tilted-{Rowland}-spectrometer on {A}rcus,'' in {\em UV, X-RAY, AND GAMMA-RAY SPACE INSTRUMENTATION FOR ASTRONOMY XX},  O.~Siegmund, Ed., {\em Proceedings of SPIE} {\bf 10397}, SPIE  (2017).
\newblock Conference on UV, X-Ray, and Gamma-Ray Space Instrumentation for Astronomy XX, San Diego, CA, AUG 06-08, 2017.

\bibitem{moritzAPJ2024}
H.~M. G\"unther, C.~DeRoo, R.~K. Heilmann, {\em et~al.}, ``Concept of a double tilted {Rowland} spectrograph for x-rays.'' submitted to ApJ  (2024).

\bibitem{SPIE2018}
R.~K. Heilmann, A.~R. Bruccoleri, J.~Song, {\em et~al.}, ``Blazed transmission grating technology development for the {A}rcus x-ray spectrometer {Explorer},'' in {\em SPACE TELESCOPES AND INSTRUMENTATION 2018: ULTRAVIOLET TO GAMMA RAY},  J.~DenHerder, S.~Nikzad, and K.~Nakazawa, Eds., {\em Proceedings of SPIE} {\bf 10699}, SPIE  (2018).
\newblock Conference on Space Telescopes and Instrumentation - Ultraviolet to Gamma Ray, Austin, TX, JUN 10-15, 2018.

\bibitem{cxc}
C.~Canizares, J.~Davis, D.~Dewey, {\em et~al.}, ``The {Chandra} high-energy transmission grating: Design, fabrication, ground calibration, and 5 years in flight,'' {\em PUBLICATIONS OF THE ASTRONOMICAL SOCIETY OF THE PACIFIC} {\bf 117}, 1144--1171  (2005).

\bibitem{RGS}
J.~den Herder, A.~Brinkman, S.~Kahn, {\em et~al.}, ``The reflection grating spectrometer on board {XMM-Newton},'' {\em ASTRONOMY \& ASTROPHYSICS} {\bf 365}, L7--L17  (2001).

\bibitem{moritz_SPIE2016}
H.~M. G\"uenther, M.~W. Bautz, R.~K. Heilmann, {\em et~al.}, ``Ray-tracing critical-angle transmission gratings for the {X-ray Surveyor} and {Explorer}-size missions,'' in {\em SPACE TELESCOPES AND INSTRUMENTATION 2016: ULTRAVIOLET TO GAMMA RAY},  J.~DenHerder, T.~Takahashi, and M.~Bautz, Eds., {\em Proceedings of SPIE} {\bf 9905}(1), SPIE  (2016).
\newblock Conference on Space Telescopes and Instrumentation - Ultraviolet to Gamma Ray, Edinburgh, SCOTLAND, JUN 26-JUL 01, 2016.

\bibitem{moritz_SPIE2018}
H.~M. G\"unther, C.~DeRoo, R.~K. Heilmann, {\em et~al.}, ``Ray-tracing {A}rcus in {Phase A},'' in {\em SPACE TELESCOPES AND INSTRUMENTATION 2018: ULTRAVIOLET TO GAMMA RAY},  J.~DenHerder, S.~Nikzad, and K.~Nakazawa, Eds., {\em Proceedings of SPIE} {\bf 10699}, SPIE  (2018).
\newblock Conference on Space Telescopes and Instrumentation - Ultraviolet to Gamma Ray, Austin, TX, JUN 10-15, 2018.

\bibitem{Alex_JVST2013}
A.~Bruccoleri, D.~Guan, P.~Mukherjee, {\em et~al.}, ``Potassium hydroxide polishing of nanoscale deep reactive-ion etched ultrahigh aspect ratio gratings,'' {\em JOURNAL OF VACUUM SCIENCE \& TECHNOLOGY B} {\bf 31}  (2013).

\bibitem{SPIE2021}
R.~K. Heilmann, A.~R. Bruccoleri, J.~Song, {\em et~al.}, ``Manufacture and performance of blazed soft x-ray transmission gratings for {A}rcus and {L}ynx,'' in {\em OPTICS FOR EUV, X-RAY, AND GAMMA-RAY ASTRONOMY X},  S.~ODell, J.~Gaskin, and G.~Pareschi, Eds., {\em Proceedings of SPIE} {\bf 11822}, SPIE  (2021).
\newblock Conference on Optics for EUV, X-Ray, and Gamma-Ray Astronomy X, San Diego, CA, AUG 01-05, 2021.

\bibitem{Alex_SPIE2013}
A.~R. Bruccoleri, D.~Guan, R.~K. Heilmann, {\em et~al.}, ``Nanofabrication advances for high efficiency critical-angle transmission gratings,'' in {\em OPTICS FOR EUV, X-RAY, AND GAMMA-RAY ASTRONOMY VI},  S.~ODell and G.~Pareschi, Eds., {\em Proceedings of SPIE} {\bf 8861}, SPIE  (2013).
\newblock Conference on Optics for EUV, X-Ray, and Gamma-Ray Astronomy VI as part of the SPIE Optics + Photonics International Symposium on Optical Engineering + Applications, San Diego, CA, AUG 26-29, 2013.

\bibitem{EIPBN2016}
A.~R. Bruccoleri, R.~K. Heilmann, and M.~L. Schattenburg, ``Fabrication process for 200 nm-pitch polished freestanding ultrahigh aspect ratio gratings,'' {\em JOURNAL OF VACUUM SCIENCE \& TECHNOLOGY B} {\bf 34}  (2016).

\bibitem{SPIE2020}
R.~K. Heilmann, A.~R. Bruccoleri, J.~Song, {\em et~al.}, ``Toward volume manufacturing of high-performance soft x-ray critical-angle transmission gratings,'' in {\em SPACE TELESCOPES AND INSTRUMENTATION 2020: ULTRAVIOLET TO GAMMA RAY},  J.~DenHerder, S.~Nikzad, and K.~Nakazawa, Eds., {\em Proceedings of SPIE} {\bf 11444}, SPIE  (2021).
\newblock Conference on Space Telescopes and Instrumentation - Ultraviolet to Gamma Ray / SPIE Astronomical Telescopes + Instrumentation Conference, DEC 14-18, 2020.

\bibitem{OE2008}
R.~K. Heilmann, M.~Ahn, E.~M. Gullikson, {\em et~al.}, ``Blazed high-efficiency x-ray diffraction via transmission through arrays of nanometer-scale mirrors,'' {\em OPTICS EXPRESS} {\bf 16}, 8658--8669  (2008).

\bibitem{SPIE2008}
R.~K. Heilmann, M.~Ahn, and M.~L. Schattenburg, ``Fabrication and performance of blazed transmission gratings for x-ray astronomy,'' in {\em SPACE TELESCOPES AND INSTRUMENTATION 2008: ULTRAVIOLET TO GAMMA RAY, PTS 1 AND 2},  M.~Turner and K.~Flanagan, Eds., {\em PROCEEDINGS OF THE SOCIETY OF PHOTO-OPTICAL INSTRUMENTATION ENGINEERS (SPIE)} {\bf 7011}(1-2), 1106, SPIE; SPIE Europe  (2008).
\newblock Conference on Space Telescopes and Instrumentation 2008 - Ultraviolet to Gamma Ray, Marseille, FRANCE, JUN 23-28, 2008.

\bibitem{RCWA}
M.~Moharam, D.~Pommet, E.~Grann, {\em et~al.}, ``Stable implementation of the rigorous coupled-wave analysis for surface-relief gratings - enhanced transmittance matrix approach,'' {\em JOURNAL OF THE OPTICAL SOCIETY OF AMERICA A-OPTICS IMAGE SCIENCE AND VISION} {\bf 12}, 1077--1086  (1995).
\newblock Optical-Society-of-America 2nd Topical Meeting on Diffractive Optics, ROCHESTER, NY, JUN 06-09, 1994.

\bibitem{moritz_SPIE2023}
H.~M. G\"unther, ``{Ray-tracing {A}rcus for performance and alignment tolerances},'' in {\em UV, X-Ray, and Gamma-Ray Space Instrumentation for Astronomy XXIII},   {\bf 12678}, International Society for Optics and Photonics, SPIE  (2023).

\bibitem{SPIE2017}
R.~K. Heilmann, A.~R. Bruccoleri, J.~Song, {\em et~al.}, ``Critical-angle transmission grating technology development for high resolving power soft x-ray spectrometers on {A}rcus and {L}ynx,'' in {\em OPTICS FOR EUV, X-RAY, AND GAMMA-RAY ASTRONOMY VIII},  S.~ODell and G.~Pareschi, Eds., {\em Proceedings of SPIE} {\bf 10399}, SPIE  (2017).
\newblock Conference on Optics for EUV, X-Ray, and Gamma-Ray Astronomy VIII, San Diego, CA, AUG 08-10, 2017.

\bibitem{SPIE2015}
R.~K. Heilmann, A.~R. Bruccoleri, and M.~L. Schattenburg, ``High-efficiency blazed transmission gratings for high-resolution soft x-ray spectroscopy,'' in {\em OPTICS FOR EUV, X-RAY, AND GAMMA-RAY ASTRONOMY VII},  S.~ODell and G.~Pareschi, Eds., {\em Proceedings of SPIE} {\bf 9603}, SPIE  (2015).
\newblock Conference on Optics for EUV, X-Ray, and Gamma-Ray Astronomy VII as part of the SPIE Optics + Photonics International Symposium on Optical Engineering + Applications, San Diego, CA, AUG 10-13, 2015.

\bibitem{SPIE2016}
R.~K. Heilmann, A.~R. Bruccoleri, J.~Kolodziejczak, {\em et~al.}, ``Critical-angle x-ray transmission grating spectrometer with extended bandpass and resolving power $>$ 10,000,'' in {\em SPACE TELESCOPES AND INSTRUMENTATION 2016: ULTRAVIOLET TO GAMMA RAY},  J.~DenHerder, T.~Takahashi, and M.~Bautz, Eds., {\em Proceedings of SPIE} {\bf 9905}, SPIE  (2016).
\newblock Conference on Space Telescopes and Instrumentation - Ultraviolet to Gamma Ray, Edinburgh, SCOTLAND, JUN 26-JUL 01, 2016.

\bibitem{ApJ_2022}
R.~K. Heilmann, A.~R. Bruccoleri, V.~Burwitz, {\em et~al.}, ``X-ray performance of critical-angle transmission grating prototypes for the {Arcus} mission,'' {\em ASTROPHYSICAL JOURNAL} {\bf 934}  (2022).

\bibitem{AO2019}
R.~K. Heilmann, J.~Kolodziejczak, A.~R. Bruccoleri, {\em et~al.}, ``Demonstration of resolving power $\lambda /{\Delta}\lambda > 10{,}000$ for a space-based x-ray transmission grating spectrometer,'' {\em APPLIED OPTICS} {\bf 58}, 1223--1238  (2019).

\bibitem{marshall:2012}
H.~L. {Marshall}, ``{Updating the Chandra HETGS efficiencies using in-orbit observations},'' in {\em Space Telescopes and Instrumentation 2012: Ultraviolet to Gamma Ray},  T.~{Takahashi}, S.~S. {Murray}, and J.-W.~A. {den Herder}, Eds., {\em Society of Photo-Optical Instrumentation Engineers (SPIE) Conference Series} {\bf 8443}, 844348  (2012).

\bibitem{SPIE2023}
R.~K. Heilmann, A.~R. Bruccoleri, E.~M. Gullikson, {\em et~al.}, ``{Soft x-ray performance and fabrication of flight-like blazed transmission gratings for the x-ray spectrometer on Arcus Probe},'' in {\em Optics for EUV, X-Ray, and Gamma-Ray Astronomy XI},  S.~L. O'Dell, J.~A. Gaskin, G.~Pareschi, {\em et~al.}, Eds.,  {\bf 12679}, 126790L, International Society for Optics and Photonics, SPIE  (2023).

\bibitem{Predehl_1992}
P.~Predehl, H.~Kraus, H.~W. Braeuninger, {\em et~al.}, ``Grating elements for the {AXAF} low-energy transmission grating spectrometer,'' in {\em {EUV}, {X}-{Ray}, and {Gamma}-{Ray} {Instrumentation} for {Astronomy} {III}},   {\bf 1743}, 475--481, SPIE  (1992).

\bibitem{Resolve2022}
Y.~Ishisaki, R.~L. Kelley, H.~Awaki, {\em et~al.}, ``{Status of Resolve instrument onboard X-Ray Imaging and Spectroscopy Mission (XRISM)},'' in {\em Space Telescopes and Instrumentation 2022: Ultraviolet to Gamma Ray},  J.-W.~A. den Herder, S.~Nikzad, and K.~Nakazawa, Eds.,  {\bf 12181}, 121811S, International Society for Optics and Photonics, SPIE  (2022).

\bibitem{Song_SPIE2018}
J.~Song, R.~K. Heilmann, A.~R. Bruccoleri, {\em et~al.}, ``Metrology for quality control and alignment of {CAT} grating spectrometers,'' in {\em SPACE TELESCOPES AND INSTRUMENTATION 2018: ULTRAVIOLET TO GAMMA RAY},  J.~DenHerder, S.~Nikzad, and K.~Nakazawa, Eds., {\em Proceedings of SPIE} {\bf 10699}, SPIE  (2018).
\newblock Conference on Space Telescopes and Instrumentation - Ultraviolet to Gamma Ray, Austin, TX, JUN 10-15, 2018.

\bibitem{JungkiEIPBN}
J.~Song, R.~K. Heilmann, A.~R. Bruccoleri, {\em et~al.}, ``Characterizing profile tilt of nanoscale deep-etched gratings via x-ray diffraction,'' {\em JOURNAL OF VACUUM SCIENCE \& TECHNOLOGY B} {\bf 37}  (2019).

\bibitem{SPIE2022}
R.~K. Heilmann, A.~R. Bruccoleri, V.~Burwitz, {\em et~al.}, ``{Flight-like critical-angle transmission grating x-ray performance for Arcus},'' in {\em Space Telescopes and Instrumentation 2022: Ultraviolet to Gamma Ray},  J.-W.~A. den Herder, S.~Nikzad, and K.~Nakazawa, Eds.,  {\bf 12181}, 1218116, International Society for Optics and Photonics, SPIE  (2022).

\bibitem{SPIE2019}
R.~K. Heilmann, A.~R. Bruccoleri, J.~Song, {\em et~al.}, ``Progress in x-ray critical-angle transmission grating technology development,'' in {\em OPTICS FOR EUV, X-RAY, AND GAMMA-RAY ASTRONOMY IX},  S.~ODell and G.~Pareschi, Eds., {\em Proceedings of SPIE} {\bf 11119}, SPIE  (2019).
\newblock Conference on Optics for EUV, X-Ray, and Gamma-Ray Astronomy IX as part of the SPIE Optics + Photonics International Symposium on Optical Engineering + Applications, San Diego, CA, AUG 13-15, 2019.

\bibitem{Gaskin_JATIS2019}
J.~A. Gaskin, D.~A. Swartz, A.~Vikhlinin, {\em et~al.}, ``Lynx x-ray observatory: an overview,'' {\em JOURNAL OF ASTRONOMICAL TELESCOPES INSTRUMENTS AND SYSTEMS} {\bf 5}  (2019).

\bibitem{moritz_JATIS2019}
H.~M. G\"uenther and R.~K. Heilmann, ``Lynx soft x-ray critical-angle transmission grating spectrometer,'' {\em JOURNAL OF ASTRONOMICAL TELESCOPES INSTRUMENTS AND SYSTEMS} {\bf 5}  (2019).

\bibitem{CheimetsJATIS2024}
P.~Cheimets and H.~M. G\"unther, ``Arcus co-alignment procedure and requirements.'' submitted to JOURNAL OF ASTRONOMICAL TELESCOPES INSTRUMENTS AND SYSTEMS  (2024).

\bibitem{FranceJATIS2024}
K.~France, ``Arcus {Probe} {UVS} instrument.'' submitted to JOURNAL OF ASTRONOMICAL TELESCOPES INSTRUMENTS AND SYSTEMS  (2024).

\end{thebibliography}
\bibliographystyle{spiejour}   


\vspace{2ex}\noindent\textbf{Ralf K. Heilmann} is a principal research scientist at the MIT Kavli Institute for Astrophysics and Space Research, and the associate director of the Space Nanotechnology Laboratory (SNL). He received his Diplom in physics from the FAU Erlangen/N\"urnberg (1991), and his MS (1993) and PhD (1996) in physics from Carnegie Mellon University. After a postdoc at Harvard he joined the SNL and has since focused on advanced lithography and the development of x-ray optics for astronomy.  He is a Senior Member of SPIE.

\vspace{2ex}\noindent\textbf{Hans Moritz G\"unther} is a research scientist at MIT. He received his undergraduate degree (in
2005) and his PhD (in 2009) in physics from the University of Hamburg, Germany. After that,
he worked at the Harvard-Smithsonian Center for Astrophysics and came to MIT in 2015. He
is currently the lead developer of MARX, the ray-tracing software used for the Chandra X-ray
observatory. His science interests are in star formation using data from the radio to X-rays.

\vspace{2ex}\noindent\textbf{Mark L. Schattenburg} is a senior research scientist at the MIT Kavli Institute for Astrophysics and Space Research, and director of the Space Nanotechnology Laboratory (SNL).  He received his Ph.D. in physics from MIT (1984). He is a Fellow of Optica.

\vspace{1ex}
\noindent Biographies and photographs of the other authors are not available.

\listoffigures
\listoftables

\end{spacing}
\end{document}